\documentclass[english]{article}
\usepackage[T1]{fontenc}
\usepackage[latin9]{inputenc}
\usepackage{geometry}
\usepackage{float}
\usepackage{amsmath}
\usepackage{amssymb}
\usepackage{graphicx}
\usepackage{setspace}
\makeatletter
\date{}

\makeatother

\usepackage{babel}

\usepackage[svgnames]{xcolor}
\usepackage{xstring}
\newcommand{\EatOneArg}[1]{}

\begin{document}
\global\long\def\dyad#1{\underline{\underline{\boldsymbol{#1}}}}%
\global\long\def\ubar#1{\underbar{\ensuremath{\boldsymbol{#1}}}}%
\global\long\def\integer{\mathbb{Z}}%
\global\long\def\natural{\mathbb{N}}%
\global\long\def\real#1{\mathbb{R}^{#1}}%
\global\long\def\complex#1{\mathbb{C}^{#1}}%
\global\long\def\defined{\triangleq}%
\global\long\def\trace{\text{trace}}%
\global\long\def\del{\nabla}%
\global\long\def\cross{\times}%
\global\long\def\diff#1#2{\frac{\partial#1}{\partial#2}}%
\global\long\def\Diff#1#2{\frac{d#1}{d#2}}%
\global\long\def\bra#1{\left\langle #1\right|}%
\global\long\def\ket#1{\left|#1\right\rangle }%
\global\long\def\braket#1#2{\left\langle #1|#2\right\rangle }%
\global\long\def\ketbra#1#2{\left|#1\right\rangle \left\langle #2\right|}%
\global\long\def\identity{\mathbf{1}}%
\global\long\def\paulix{\begin{pmatrix}  &  1\\
 1 
\end{pmatrix}}%
\global\long\def\pauliy{\begin{pmatrix}  &  -i\\
 i 
\end{pmatrix}}%
\global\long\def\pauliz{\begin{pmatrix}1\\
  &  -1 
\end{pmatrix}}%
\global\long\def\sinc{\mbox{sinc}}%
\global\long\def\four{\mathcal{F}}%
\global\long\def\dag{^{\dagger}}%
\global\long\def\norm#1{\left\Vert #1\right\Vert }%
\global\long\def\hamil{\mathcal{H}}%
\global\long\def\tens{\otimes}%
\global\long\def\ord#1{\mathcal{O}\left(#1\right)}%
\global\long\def\undercom#1#2{\underset{_{#2}}{\underbrace{#1}}}%
 
\global\long\def\conv#1#2{\underset{_{#1\rightarrow#2}}{\longrightarrow}}%
\global\long\def\tg{^{\prime}}%
\global\long\def\ttg{^{\prime\prime}}%
\global\long\def\clop#1{\left[#1\right)}%
\global\long\def\opcl#1{\left(#1\right]}%
\global\long\def\broket#1#2#3{\bra{#1}#2\ket{#3}}%
\global\long\def\div{\del\cdot}%
\global\long\def\rot{\del\cross}%
\global\long\def\up{\uparrow}%
\global\long\def\down{\downarrow}%
\global\long\def\Tr{\mbox{Tr}}%

\global\long\def\per{\mbox{}}%
\global\long\def\pd{\mbox{}}%
\global\long\def\p{\mbox{}}%
\global\long\def\ad{\mbox{}}%
\global\long\def\a{\mbox{}}%
\global\long\def\la{\mbox{\ensuremath{\mathcal{L}}}}%
\global\long\def\cm{\mathcal{M}}%
\global\long\def\cg{\mbox{\ensuremath{\mathcal{G}}}}%

\title{Kinetic equation for weak interaction of directional internal waves}
\author{Michal Shavit, Oliver B\"uhler 
and Jalal Shatah\\
\emph{\normalsize{}Courant Institute of Mathematical Sciences, New
York University, NY 10012, USA }}
\maketitle
\begin{abstract}
Starting from the two dimensional Boussinesq equation without rotation
we derive a kinetic equation for weak interaction of internal waves
using non canonical variables. We follow a formalism introduced by
P. Ripa in the 80's. The advantage of this formalism
is that it describes the system in terms of the natural linear eigenfunctions
of eastward and westward propagating internal waves. Using properties
of orthogonality of the eigenfunctions with respect to a (pseudo) metric set
by the energy we can write non perturbative theory for the interaction
of waves given in terms of the expansion amplitudes. The evolution
is controlled by a system of equations, with quadratic nonlinearity,
which is an exact representation of the original model equations.
The dynamics is constrained by the conservation of energy and pseudo-momentum,
which can be written simply as a linear combination of the squared
absolute value of the amplitudes. The possibility of a generalization
of the Fjortoft\textquoteright s argument to internal gravity waves
and observation of a non trivial double cascade of energy and pseudo-momentum
is discussed. 
\end{abstract}
\tableofcontents{}

\section{{\large{}Introduction}}
Consider the equations for a three dimensional incompressible fluid
with a vertically stratified density within the Boussinesq approximation: 
\begin{align}
\left(\partial_{t}+\mathbf{v}\cdot\del\right)\mathbf{v}+f\hat{z}\cross\mathbf{v} & =-\frac{\del p}{\rho_{r}}+b\hat{z}\\
\left(\partial_{t}+\mathbf{v}\cdot\del\right)b+N^{2}\mathbf{v}\cdot\hat{z} & =0\\
\div\mathbf{v} & =0
\end{align}
where $\mathbf{v}:\Omega^{3}\rightarrow\real 3$ is the velocity field,
$p$ is the pressure deviation from the hydrostatic pressure distribution,
$\rho_{r}$ a constant reference density, $b=\left(\rho_{0}\left(z\right)-\rho\right)g/\rho_{r}$
is the buoyancy field. Here, $\rho$ is the total density field, $\rho_{0}\left(z\right)$
is the background density profile and $-g\hat{z}$ the gravity force.
For the validity of the Boussinesq approximation the density difference
is assumed to satisfy $\left|\rho-\rho_{r}\right|/\rho_{r}\ll1$.
The buoyancy frequency $N=\sqrt{\left(-g/\rho_{r}\right)\left(\partial_{z}\rho_{0}\right)}$
is taken to be constant and $f\hat{z}\cross\mathbf{v}$ is the Coriolis
force within the $f$-plane approximation. The domain can be periodic
$\Omega=\mathbb{T}_{L}:=\left[0,L\right]$ with periodic boundary
conditions or $\Omega=\real{}$. To keep the notation simple, we consider
$\Omega=\real{}.$ Neglecting earth rotation we reduce the system
to two dimensions considering one horizontal direction $x$ and one
vertical $z$. Following a formalism introduced by \cite{Ripa}, we
rewrite the equations in the following form
\begin{align}
\left(\partial_{t}+\mathbf{v}\cdot\del\right)D\phi-L\phi & =0\label{eq:Bous}
\end{align}
where
\begin{align}
D & =\begin{pmatrix}-\Delta & 0\\
0 & N^{2}
\end{pmatrix},\,\,\,L=N^{2}\begin{pmatrix}0 & 1\\
1 & 0
\end{pmatrix}\partial_{x}.
\end{align}
and
\begin{align}
\phi & =\begin{pmatrix}A\\
\zeta
\end{pmatrix}
\end{align}
is a two dimensional vector that consists of the stream function $A$,
so that $\mathbf{v}=\left(-\partial_{z},\partial_{x}\right)A$, and
the vertical displacement $\zeta=-\frac{b}{N^{2}}$. Note that the
two entries of $\phi$ carry different dimensions. The Boussinesq
system in two dimensions has two integrals of motion, the energy $E=\frac{1}{2}\int d\mathbf{x}\left((\del A)^2+\zeta^{2}\right)$
and pseudo-momentum (PM) $P=\int d\mathbf{x}\zeta\Delta A$. We expand
the field $\phi$
\begin{align}
\phi & =\int\!\! d\alpha z_{\alpha}e_{\alpha},\label{eq:expansion}
\end{align}
in terms of solutions to the eigenvalue problem that corresponds to
the linear part of (\ref{eq:Bous})
\begin{align}
L e_{\alpha} & =-i\omega_\alpha D e_{\alpha}.
\end{align}
The eigenmodes are 
\begin{align}
e_{\alpha} & =M^{-1}\left(K\right)\begin{pmatrix}-1\\
s_{\alpha}
\end{pmatrix}e^{i\mathbf{k}\cdot\mathbf{x}}
\end{align}
where $s_{\alpha}=k_{x}\omega_{\alpha}^{-1}$ is the slowness, 
$M\left(K\right)$ is a normalization and $K=\sqrt{\boldsymbol{k\cdot k}}$ is the wave number absolute value. Waves that carry positive
(negative) slowness correspond to eastward (westward) propagating
waves. $D$ is hermitian and semipositive definite and $L$ is skew hermitian
which makes the eigenvalues
\begin{align}
\omega_{\alpha} & =\omega\left(\sigma,\mathbf{k}\right)=\sigma N k_x/K \end{align}
real and the eigenvectors corresponding to different eigenvalues $D$-orthogonal.
The three index $\alpha=\left(\sigma=\pm1,\mathbf{k}\right)$ indicates
the frequency branch and wave number, so by the integral in (\ref{eq:expansion})
we mean $\int\!\! d\alpha=\int\!\! d\sigma \int\!\!d\mathbf{k}$ and  $ d \sigma=\sum_{\sigma=\pm1}$ is the counting measure. Note that $D$ defines a seminorm in the space spanned by $\left\{ e_{\alpha}\right\} $
in terms of the energy
\begin{align}
E & =\frac{1}{2}\int\!\! d\mathbf{x}\phi^{\dagger}D\phi,\label{eq:e metric}
\end{align}
so we pick the normalization $M=\sqrt{2}K$
s.t eigenvectors corresponding to different eigenvalues are $D$-orthonormal
\begin{align}
\int\!\! e_{\beta}^{\dagger}D e_{\alpha} d\boldsymbol{x}  =\delta_{\sigma_{\alpha}\sigma_{\beta}}\delta\left(\mathbf{k}_{\alpha}-\mathbf{k}_{\beta}\right).
\end{align}
When writing (\ref{eq:e metric}) we acknowledge the subtlety of infinite integration. Resolving infinities that can arise from integration over an unbounded domain is deferred to future work. For the purpose of this text, it is assumed that all integrals are locally bounded.
Since $\phi$ is real the fields $z_{\alpha}$ satisfy
\begin{align}
z_{-}\left(\mathbf{k}\right) & =z_{-}^{*}\left(-\mathbf{k}\right)\label{eq:reality}\\
z_{+}\left(\mathbf{k}\right) & =z_{+}^{*}\left(-\mathbf{k}\right)
\end{align}
so that if we separate integration over the two branches
\begin{align}
\phi & =\int\! d\mathbf{k}z_{+} e_{+}\left(\mathbf{k}\right)+\int\! d\mathbf{k}z_{-} e_{-}\left(\mathbf{k}\right)=\phi_{+}+\phi_{-},
\end{align}
both components $\phi_{+}$ and $\phi_{-}$ are real functions. In terms of the expansion amplitudes, 
$z_{\alpha}$, the equations of motion are
\begin{align}
z_{\alpha,t}+i\omega_{\alpha}z_{\alpha} & =\frac{1}{2} \int\! d\beta d\gamma\ \sigma_{\alpha}^{\beta\gamma}z_{\beta}^{*}z_{\gamma}^{*},\label{eq:z}
\end{align}
where $\sigma_{\alpha}^{\beta\gamma}$ are the interaction coefficients
given by
\begin{align}
\sigma_{\alpha}^{\beta\gamma} & =- \int\! d\mathbf{x}\left(\hat{\phi}_{\alpha}^{\dagger}\left(\mathbf{v}_{\beta}\left(k,m\right)\cdot\del D\hat{\phi}_{\gamma}+\mathbf{v}_{\gamma}\left(k,m\right)\cdot\del D\hat{\phi}_{\beta}\right)^{*}\right)\\
 & =-N^{2}M_{\alpha}^{-1}M_{\beta}^{-1}M_{\gamma}^{-1}\mathbf{k}_{\beta}\cross\mathbf{k}_{\gamma}\left(s_{\beta}+s_{\gamma}+s_{\alpha}\right)\left(s_{\gamma}-s_{\beta}\right)\left(2\pi\right)^{2}\delta\left(\mathbf{k}_{\alpha,\beta,\gamma}\right).
\end{align}
From the last line it is clear that waves with the same slowness
$s_{\gamma}=s_{\beta}$ or with wave numbers that are linearly dependent
$\mathbf{k}_{\gamma}=a\mathbf{k}_{\beta}$ do not interact and are
exact solutions of the nonlinear equations. The interaction of a triad
with $s_{\beta}+s_{\gamma}+s_{\alpha}=0$ vanishes as well. The integrals
of motion are both quadratic in terms of the expansion amplitudes and are given simply by the linear combinations
\begin{align}
E & =\frac{1}{2}\int\! d\alpha z_{\alpha}z_{\alpha}^{*},\label{eq:energy z}\\
P & =\frac{1}{2}\int\! d\alpha s_{\alpha}z_{\alpha}z_{\alpha}^{*}.\label{eq:pm z}
\end{align}
where (\ref{eq:energy z}) follows from (\ref{eq:e metric}), and (\ref{eq:pm z})
from the reality conditions (\ref{eq:reality}). Due to conservation
of energy and PM the interaction coefficients satisfy
\begin{align}
\sigma_{\alpha}^{\beta\gamma}+\sigma_{\beta}^{\alpha\gamma}+\sigma_{\gamma}^{\beta\alpha} & =0,\label{eq:energy co}\\
s_{\alpha}\sigma_{\alpha}^{\beta\gamma}+s_{\beta}\sigma_{\beta}^{\alpha\gamma}+s_{\gamma}\sigma_{\gamma}^{\beta\alpha} & =0.\label{eq:PM co}
\end{align}
The last two equations can be readily obtained by writing the time
derivative of the energy in terms of the expansion amplitudes $z_{\alpha}$: 
\begin{align}
0=\dot{E} & =\frac{1}{2}\int\! d\alpha \dot{E}_{\alpha}=\frac{1}{4}\int\! d\alpha d\beta d\gamma\ \sigma_{\alpha}^{\beta\gamma}\Re\left(z_{\beta}^{*}z_{\gamma}^{*}z_{\alpha}^{*}\right)
\end{align}
for each triad $(\alpha,\beta,\gamma)$ the monomial $\Re\left(z_{\beta}^{*}z_{\gamma}^{*}z_{\alpha}^{*}\right)$
in the sum above appears with the coefficients $\left(\sigma_{\alpha}^{\beta\gamma}+\sigma_{\gamma}^{\beta\alpha}+\sigma_{\beta}^{\alpha\gamma}\right)$,
so that energy is conserved if and only if (\ref{eq:energy co}) is
satisfied, similarly for the time derivative of the PM. These relations
can be also verified directly. Note that equations (\ref{eq:z},\ref{eq:energy co},\ref{eq:PM co})
describe the Boussinesq system in the form of a hydrodynamic type system
with two integrals of motion \cite{Obukhov,hydro two}. The plan of
the text is as follows: in section \ref{sec:Kinetic-equation-for} we derive
a kinetic equation from (\ref{eq:z}) for weakly interacting internal
waves, discuss its equipartition solutions and comment on constant
flux solutions. In section \ref{sec:Double-cascades-and} we describe the
motivation for studying a truncated kinetic equation that includes
only interactions of eastward propagating waves, the possibility of
a generalization of the Fjortoft\textquoteright s argument to internal
waves and observation of a non trivial double cascade of energy and
PM. 

\section{{\large{}Kinetic equation for eastward and westward propagating internal
waves \label{sec:Kinetic-equation-for}}}

\begin{figure}[H]
\centering{}\includegraphics[scale=0.4]{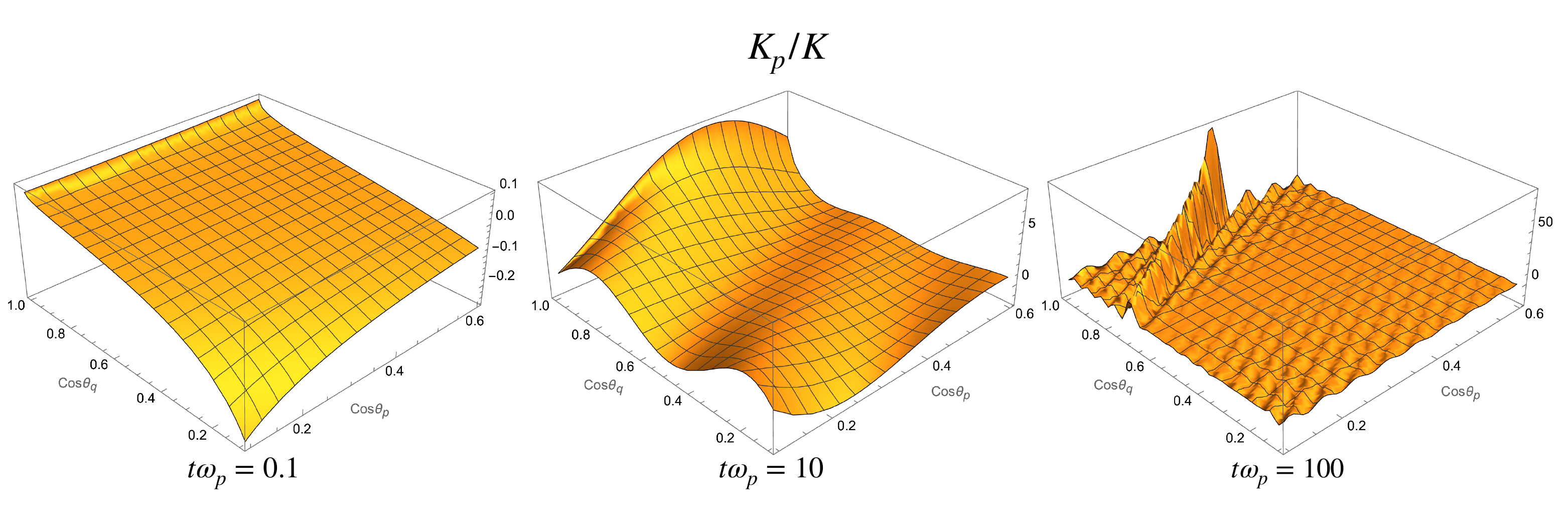}\caption{{\small{}Concentration of the scaled wave number amplitude of eastward
propagating interacting waves on the resonant manifold. For three
interacting eastward propagating waves with wave numbers that satisfy
$\mathbf{q}+\mathbf{p}+\mathbf{k}=0$, with $\mathbf{p}=K_{p}\left(\cos\theta_{p},\sin\theta_{p}\right)$,
the ratio of the absolute values of two wave numbers is plotted as a function
of the wave number angles, $\cos\theta_{p}$ and $\cos\theta_{q}$.
As $t\omega_{p}\gg1$ the possible values for interaction concentrate
on the one dimensional resonant manifold.}}
\end{figure}

Assume that non linear interaction is small compared to the linear evolution, so that we can write the equations of motion (\ref{eq:z}) introducing a small uniform parameter $\epsilon$,
\begin{align}
z_{\alpha,t}+i\omega_{\alpha}z_{\alpha} & =\frac{1}{2} \epsilon \int\!\! d\beta d\gamma \sigma_{\alpha}^{\beta\gamma}z_{\beta}^{*}z_{\gamma}^{*}.\label{eq:z_epsilon_a}
\end{align}
Such small uniform parameter relevant
for internal wave interaction is the root mean square vertical gradient
of the vertical displacement
\begin{align}
\epsilon & =\sqrt{\left\langle \left(\partial_{z}\zeta\right)^{2}\right\rangle }.
\end{align}
The discussion can also be generalized to a non uniform small parameter $\epsilon=\epsilon_{\alpha}$.
Transforming to envelopes $z_{\alpha}\rightarrow z_{\alpha}e^{-i\omega_{\alpha}t}$
(\ref{eq:z_epsilon_a}) becomes, 
\begin{align}
\dot{z_{\alpha}} & =\frac{1}{2}\epsilon\int\!\! d\beta d\gamma \sigma_{\alpha}^{\beta\gamma}z_{\beta}^{*}z_{\gamma}^{*}e^{i\Omega_{\alpha\beta\gamma}t}.\label{eq:z epsilon}
\end{align}
where $\Omega_{\alpha\beta\gamma}=\omega_{\gamma}+\omega_{\beta}+\omega_{\alpha}$
is the sum of the linear frequencies of the interacting waves. Assuming
initial conditions are random, we are interested in writing an evolution
equation up to order $\epsilon^{2}$ to the averaged energy density
\begin{align}
n_{\alpha}:& =\left\langle z_{\alpha}z_{\alpha}^{*}\right\rangle ,
\end{align}
where $\left\langle \cdot\right\rangle$ denotes the average with
respect to the initial data distribution. In order to write the kinetic equation we expand
the amplitudes $z_{\alpha}$ in terms of the initial data using integration
by parts. Integrating wrt time (\ref{eq:z epsilon}), we write
\begin{align}
z_{\alpha}\left(t\right) & =z_{\alpha}\left(0\right)+\frac{1}{2}\epsilon\int_{0}^{t}\! ds\int\!\! d\beta d\gamma \sigma_{\alpha}^{\beta\gamma}z_{\beta}^{*}z_{\gamma}^{*}e^{i\Omega_{\alpha\beta\gamma}s}.
\end{align}
One integration by parts yields
\begin{align}
z_{\alpha}\left(t\right) &  =z_{\alpha}\left(0\right)+\frac{1}{2}\epsilon\int\!\! d\beta d\gamma  \sigma_{\alpha}^{\beta\gamma}z_{\beta}^{*}\left(0\right)z_{\gamma}^{*}\left(0\right)G_{1}\left(t,0\right)+\frac{1}{2}\epsilon\int_{0}^{t}\!ds
\int\!\! d\beta d\gamma \sigma_{\alpha}^{\beta\gamma}\frac{d}{ds}\left(z_{\beta}^{*}z_{\gamma}^{*}\right)G_{1}\left(t,s\right)
\end{align}
where
\begin{align}
G_{1}\left(t,s\right) & =\int_{s}^{t}e^{i\Omega_{\alpha\beta\gamma}\tau}d\tau,
\end{align}
is chosen so that $G_{1}\left(t,t\right)=0$. In
order to close the kinetic equation at order $\epsilon^{2}$ we need
to integrate by parts one more time
\begin{align}
z_{\alpha}\left(t\right) & =z_{\alpha}\left(0\right)+\frac{1}{2}\epsilon\int\!\! d\beta d\gamma \sigma_{\alpha}^{\beta\gamma}z_{\beta}^{*}z_{\gamma}^{*}\left(0\right)G_{1}\left(t,0\right)\\
 & +\frac{1}{2}\epsilon\int\!\! d\beta d\gamma  \sigma_{\alpha}^{\beta\gamma}\left(\left(\frac{d}{ds}z_{\beta}^{*}z_{\gamma}^{*}\right)\mid_{s=0}G_{2}^{\beta}\left(t,0\right)+\left(z_{\beta}^{*}\frac{d}{ds}z_{\gamma}^{*}\right)\mid_{s=0}G_{2}^{\gamma}\left(t,0\right)\right)+\frac{1}{2}\epsilon\int_{0}^{t}\! ds \int\!\! d\beta d\gamma \sigma_{\alpha}^{\beta\gamma}P_{4}\left(z\right)G_{2}\left(t,s\right)ds\nonumber,
\end{align}
where
\begin{align}
G_{2}^{\beta}\left(t,s\right) & =\int_{t}^{s}e^{-i\Omega_{\beta\beta'\gamma'}s_{1}}G_{1}\left(t,s_{1}\right)ds_{1},
\end{align}
and $P_{4}\left(z\right)$ is a term that includes fourth order polynomials
of the amplitudes and $P_{4}\left(z\right)\mid_{t=0}=\frac{d^{2}}{ds^{2}}\left(z_{\beta}^{*}z_{\gamma}^{*}\right)\mid_{t=0}$
. By the product $P_{4}\left(z\right)G_{2}\left(t,s\right)$ we mean
that each monomial in $P_{4}$ is multiplied by its corresponding
time dependent term $G_{2}^{\beta}\left(t,s\right)$. 

Assuming the initial distribution of amplitudes is exactly Gaussian,
so that the off diagonal second order correlators are zero initially: $\left\langle z_{\alpha}^{2}\right\rangle ,\left\langle z_{+}\left(\mathbf{k}\right)z_{-}\left(\mathbf{k}\right)\right\rangle ,\left\langle z_{+}\left(\mathbf{k}\right)z_{-}^{*}\left(\mathbf{k}\right)\right\rangle =0$,
we can write an equation for the average energy density: 
\begin{align}
n_{\alpha} & =\left\langle z_{\alpha}z_{\alpha}^{*}\right\rangle \left(t\right)=\left\langle z_{\alpha}z_{\alpha}^{*}\right\rangle \left(0\right)+\epsilon^{2}\int\!\! d\beta d\gamma \ \sigma_{\alpha}^{\beta\gamma}\left(\sigma_{\beta}^{\alpha\gamma}n_{\alpha}n_{\gamma}+\sigma_{\gamma}^{\alpha\beta}n_{\beta}n_{\alpha}+\sigma_{\alpha}^{\beta\gamma}n_{\beta}n_{\gamma}\right)\frac{1-\cos\Omega_{\alpha\beta\gamma}t}{\Omega_{\alpha\beta\gamma}^{2}}+O\left(\epsilon^{4}\right).\label{eq:pre kin}
\end{align}
Note that terms that include odd powers of $\epsilon$ correspond
to odd monomials of the amplitudes $z_{\alpha}$, these contributions
vanish due to the Gaussianity of the initial distribution, so the
kinetic equations includes only even power in $\epsilon$. In order to derive the kinetic equation one should start from the discrete case of a finite domain $\Omega=\left[0,L\right]$ and take the kinetic limits $L\rightarrow\infty$ and $t\omega\rightarrow\infty$ in the correct order and rate. In this process we assume that there are enough quasi-resonances wrt to exact resonances so the derivation occurs. We can then write $\lim_{t\omega\rightarrow\infty}\frac{1-\cos\omega_{\alpha\beta\gamma}t}{\omega_{\alpha\beta\gamma}^{2}}=\pi t\delta\left(\omega_{\alpha\beta\gamma}\right)$ and arrive at the kinetic equation
\begin{align}
\dot{n_{\alpha}} & =\pi\epsilon^{2}\int\!\! d\beta d\gamma \  \sigma_{\alpha}^{\beta\gamma}\left(\sigma_{\beta}^{\alpha\gamma}n_{\alpha}n_{\gamma}+\sigma_{\gamma}^{\alpha\beta}n_{\beta}n_{\alpha}+\sigma_{\alpha}^{\beta\gamma}n_{\beta}n_{\gamma}\right)\delta\left(\Omega_{\alpha\beta\gamma}\right).\label{eq:kin}
\end{align}
All off diagonal correlators except for $\left\langle z_{+}\left(\mathbf{k}\right)z_{-}^{*}\left(\mathbf{k}\right)\right\rangle $
remain zero at order $\epsilon^{2}$. The time derivative $\frac{d}{dt}\left\langle z_{+}\left(\mathbf{k}\right)z_{-}^{*}\left(\mathbf{k}\right)\right\rangle $
has $O\left(\epsilon^{2}\right)$ fluctuating contributions, so
this correlator is in general not zero pointwise; However, it averages over time weakly to zero, so that the kinetic equation (\ref{eq:kin})
remains valid. This is discussed in detail in the appendix. 

The kinetic equation (\ref{eq:kin}) has an isotropic solution that
corresponds to equipartition of energy
\begin{align}
n_{\alpha} & = T
\end{align}
as any third cumulant, i.e any summand in the collision integral of
the right hand side of (\ref{eq:kin}), is zero:
\begin{align}
\sigma_{\beta}^{\gamma\alpha}n_{\alpha}n_{\gamma}+\sigma_{\gamma}^{\beta\alpha}n_{\alpha}n_{\beta}+\sigma_{\alpha}^{\beta\gamma}n_{\gamma}n_{\beta} & =T^{2}\left(\sigma_{\beta}^{\gamma\alpha}+\sigma_{\gamma}^{\beta\alpha}+\sigma_{\alpha}^{\beta\gamma}\right)=0.
\end{align}
The last equality follows from energy conservation in the dynamical
equations (\ref{eq:energy co}). For particular cases where the slowness
is bounded from below the kinetic equation has the solution
\begin{align}
n_{\alpha}&=\frac{1}{1+T^{-1}s_{\alpha}},
\end{align}
which is isotropic, as $s_\alpha=\sigma N^{-1} K$,  and corresponds to equipartition of energy and
PM. 
The kinetic equation (\ref{eq:kin}) has acquired a second PM invariant, $P_z=\int \! d\alpha s_\alpha^zz_\alpha z_\alpha^*$, based on the vertical slowness $s_\alpha^z=k_z/\omega_\alpha$. This is not conserved in the dynamical equation (\ref{eq:z}).
 For resonantly interacting triads the ratios of the interaction coefficients
and frequencies are equal -
\begin{align}
\Gamma_{\alpha\beta\gamma}\delta\left(\mathbf{k}_{\alpha,\beta,\gamma}\right)=\frac{\sigma_{\alpha}^{\beta\gamma}}{\omega_{\alpha}}=\frac{\sigma_{\beta}^{\alpha\gamma}}{\omega_{\beta}}=\frac{\sigma_{\gamma}^{\alpha\beta}}{\omega_{\gamma}},
\end{align}
this follows from the relations among the interaction coefficients, $\sigma_{\alpha}^{\beta\gamma}$, due to energy and PM conservation: Indeed, for any resonantly interacting triad $\left(\alpha,\beta,\gamma\right)$ let $\mathbf{k}_{x}$ be the vector of the components of the wave numbers in the $\hat{x}$ direction and let $\hat{\omega}$ be the vector of the components of the frequencies. Then the three vectors $\left(\mathbf{k}_{x},\hat{\omega}-\frac{\hat{\omega}\cdot\mathbf{k}_{x}}{\mathbf{k}_{x}\cdot\mathbf{k}_{x}}\mathbf{k}_{x},\mathbf{1}\right)$, where $\mathbf{1}=\left(1,1,1\right)$, form an orthogonal basis of $\real 3$. As the vector $\frac{\sigma}{\omega}=\left(\frac{\sigma_{\alpha}^{\beta\gamma}}{\omega_{\alpha}},\frac{\sigma_{\beta}^{\alpha\gamma}}{\omega_{\beta}},\frac{\sigma_{\gamma}^{\alpha\beta}}{\omega_{\gamma}}\right)$ is perpendicular to $\mathbf{k}_{x}$ and $\hat{\omega}$, it must be proportional to $\mathbf{1}$. This brings the kinetic equation
(\ref{eq:kin}) to the canonical form 
\begin{align}
\dot{n_{\alpha}} & =\pi\epsilon^{2}\int\! d\beta d\gamma \ \omega_{\alpha}\Gamma_{\alpha\beta\gamma}^{2}\left(\omega_{\beta}n_{\alpha}n_{\gamma}+\omega_{\gamma}n_{\beta}n_{\alpha}+\omega_{\alpha}n_{\beta}n_{\gamma}\right)\delta\left(\mathbf{k}_{\alpha,\beta,\gamma}\right)\delta\left(\Omega_{\alpha\beta\gamma}\right).\label{eq:kin sim}
\end{align}
Let us stress that (\ref{eq:kin sim}) is the exact and general kinetic equation arising from the Boussinesq dynamical equations (\ref{eq:Bous}). We can write (\ref{eq:kin sim}) as a continuity equation 
\begin{align}
\dot{n}_\alpha+\text{div}\Pi\left(\mathbf{k_\alpha}\right)&=0.\label{eq:full continuity}
\end{align}
where $\Pi$ is the two dimensional (not uniquely defined) flux. Integrating (\ref{eq:kin sim},\ref{eq:full continuity}) wrt the angle $\theta_\alpha$ we write
\begin{align}
	\dot{g_\alpha}+\text{div}\Pi_{k_{\alpha}}=0\label{eq:full continuity radial},
\end{align}
where $\Pi_{k_\alpha}=K_\alpha \int_{0}^{2\pi}d\theta\Pi\cdot\hat{K}$ and $g_{\alpha}=\int_{0}^{2\pi} \! n_\alpha \left(\mathbf{k}\right)d\theta_k$ are the isotropic parts of the flux and energy spectrum, respectively. In general, (\ref{eq:full continuity radial}) can have additional solutions to (\ref{eq:full continuity}). In the case that the PM is symmetrically distributed, so that $n_{-}(k)=n_{+}(k)$ and $P=0$, (\ref{eq:full continuity radial}) has a constant flux formal solution with a Kolmogorov-Zakharov scaling: $g_k\propto K^{-(\delta+d)}$, where $\delta+d=3$ is the sum of the homogeneity degree of the interaction $\Gamma$ wrt the wave number amplitudes and dimension (see appendix for the derivation). Locality and relevance of this solution to the full kinetic equation in studied in a following paper. We note that by additionally applying the so-called "hydrostatic approximation" that brings the interaction coefficients and dispersion relations to a bi-homogeneous form
\begin{align}
\omega_{\sigma}\left(\mathbf{k}\right) & \rightarrow\sigma N^{-1} k_{x}/ \sqrt{k_{z}^{2}}
\\
\Gamma_{\alpha\beta\gamma}^{2}\rightarrow & \left(\sqrt{k_{\alpha z}^{2}}+\sqrt{k_{\beta z}^{2}}+\sqrt{k_{\gamma z}^{2}}\right)^{2},
\end{align} one can search for bi-homogeneous $n_{\alpha}\left(\mathbf{k}\right)=k_{x}^{m}k_{z}^{l}$ approximate constant flux solutions of
(\ref{eq:kin sim}) 
by applying conformal transformations similar to those suggested by
Kuznetsov \cite{Kuz,Falk}. \cite{Lvov IW} found such solutions  for a kinetic equation derived using the hydrostatic approximation for the Boussinesq equations with cylindrical symmetry in three dimensions. 
In locality analysis of the kinetic equation (\ref{eq:kin sim}) one should bear in mind that in the
inertial range relevant for the ocean, the absolute value of frequency is bounded by the buoyancy frequency and the Coriolis force  $f\leq|\omega_{\alpha}|\leq N$.

We conclude this part by mentioning that the limit $\omega_\alpha\rightarrow0 \ (k_x\rightarrow 0)$ that corresponds to pure vertical shear flows pose special problems in non-rotating 2d Boussinesq equations. One difficulty, e.g., is that the off-diagonal correlators $\left< z_{+}(k)z_{-}^*(k) \right>$ cannot be neglected in our theory. However, if no vertical shear flows exist initially, they remain zero as they cannot be created by nonlinear resonant interaction among waves since the interaction coefficients $\Gamma=\omega \sigma$ vanish in this limit. This degeneracy is removed once rotation is added.

\section{{\large{}Double cascades and wave turbulence of eastward propagating
internal waves \label{sec:Double-cascades-and}}}

The decomposition of the fields $\phi=\sum_{\alpha}z_{\alpha} e_{\alpha}$
into eastward and westward propagating waves allows us to decompose
the PM into two components, positive and negative
\begin{align}
P & =P_{+}+P_{-},
\end{align}
where we use the convention that eastward propagating waves carry
positive pseudo momentum
\begin{align}
P_{+} & =\frac{1}{2}\int_{-\infty}^{\infty}d\mathbf{k}s_{+}E_{+}\left(\mathbf{k}\right),
\end{align}
here $E_{+}=z_{+}z_{+}^{*}\left(\mathbf{k}\right)$ is the energy
density of the eastward propagating waves and remind that the positive
slowness is given by $s_{+}=k_{x}/\omega_{+}=N^{-1}K$.
The time derivatives of the positive and negative PM can be written
in terms of the time derivatives of the corresponding energy densities:
\begin{align}
\dot{P}_{\sigma=\pm} & =\frac{1}{2}\int_{-\infty}^{\infty}d\mathbf{k}s_{\sigma}\dot{E}_{\sigma}\left(\mathbf{k}\right)=\frac{1}{2}\sum_{\beta,\gamma}\int_{0}^{\infty}d\mathbf{k}s_{\sigma}\left(\mathbf{k}\right)\sigma_{\left(\sigma,\mathbf{k}\right)}^{\beta\gamma}\Re\left(z_{\beta}^{*}z_{\gamma}^{*}z_{\left(\sigma,\mathbf{k}\right)}^{*}\right).
\end{align}
Consider the initial value problem with waves solely propagating eastward,
that is $z_{-}\left(\mathbf{k},t=0\right)=0$ $\forall\mathbf{k}$. The latter
can be achieved initially if the stream function density in wave number
is proportional to the displacement by 
\begin{align}
-K\hat{\psi_{0}}\left(\mathbf{k}\right) & =N\hat{\eta_{0}}\left(\mathbf{k}\right).
\end{align}
This implies that the potential energy stored at every
wave number equals to the kinetic energy stored at this wave number.
The PM is then positive, as $P_{-}\left(t=0\right)=0$ and it's time
derivative vanishes as well $\dot{P}_{-}\left(t=0\right)=0$. That
means that for some short period of time eastward propagating
waves remain predominant in the system. This asymmetry is fixed into
the memory of the system and cannot be forgotten since the PM is an
exact integral of motion and will remain positive for all times. So every creation of a westward propagating wave due to nonlinear interaction must be accompanied by an equal-size creation of an eastward propagating wave. Creation of both positive and negative PM
waves is restricted by the conservation of both energy and PM. This has practical relevance for ocean dynamics.  For example,  strongly directional internal wave fields arise naturally in the case  of internal tides radiated away from isolated topography structures such as the Hawaiian ridge \cite{LSY03}. We are interested in studying the implications of the existence
of two positive definite invariants with proportional densities on
the dynamics and statistics of internal waves. In particular, if the Fjortoft's argument for two dimensional hydrodynamics \cite{Fjo} applies to our system as well, the power law proportionality between the energy and pseudo-momentum should lead to inverse energy transfer, from small to large scales.


One way of addressing these questions is to consider the equations truncated to the positive slowness branch:
\begin{align}
\partial_{t}z_{+}\left(\mathbf{k}\right) =\int\! d\mathbf{q}d\mathbf{p}\ \sigma_{\mathbf{k}}^{\mathbf{q}\mathbf{p}}z_{+}^{*}\left(\mathbf{p}\right)z_{+}^{*}\left(\mathbf{q}\right)e^{i\Omega_{kpq}t},\label{eq:pos z}
\end{align}
where the interaction is given by
\begin{align}
\sigma_{\mathbf{k}}^{\mathbf{q}\mathbf{p}} & =-M_{q}^{-1}M_{p}^{-1}M_{k}^{-1}\mathbf{q}\cross\mathbf{p}\left(K+K_q+K_p\right)\left(K_p-K_q\right)\delta\left(\mathbf{k}_{\alpha,\beta,\gamma}\right).
\end{align}
Theses equations have two positive quadratic integrals of motion
\begin{align}
E_{+} & =\frac{1}{2}\int d\mathbf{k}z_{+}^{*}\left(\mathbf{k}\right)z_{+}\left(\mathbf{k}\right),\\
P_{+} & =\frac{1}{2}\int d\mathbf{k}N^{-1}Kz_{+}^{*}\left(\mathbf{k}\right)z_{+}\left(\mathbf{k}\right),
\end{align}
since the following sums 
\begin{align}
\sigma_{1}^{23}+\sigma_{2}^{31}+\sigma_{3}^{23} & =0,\\
K_1\sigma_{1}^{23}+K_2\sigma_{2}^{13}+K_3\sigma_{3}^{21} & =0,
\end{align}
vanish for any triad $\mathbf{k}_{1}+\mathbf{k}_{2}+\mathbf{k}_{3}=0$,
where we used the shorthand notation $\sigma_{1}^{23}:=\sigma_{\mathbf{k}_{1}}^{\mathbf{k}_{2}\mathbf{k}_{3}}.$

We can then write a kinetic equation for the averaged energy density of the positive branch:
\begin{align}
\dot{m}\left(k\right) & =\epsilon^{2}\pi\int\! d\mathbf{q}\int\! d\mathbf{p}\omega_{k}\Gamma_{pqk}^{2}m_{k}m_{p}m_{q}\left(\omega_{q}m_{q}^{-1}+\omega_{p}m_{p}^{-1}+\omega_{k}m_{k}^{-1}\right)\delta\left(\omega_{p,q,k}\right)\delta\left(\mathbf{k}+\mathbf{q}+\mathbf{p}\right), \label{eq:pos kin}
\end{align}
where we denote $m_k=\left\langle z_{+}\left(k\right)z_{+}\left(k\right)^{*}\right\rangle$ truncated to the dynamics of (\ref{eq:pos z}). Writing the wave numbers in a polar form
$\mathbf{k} =K\left(\sin\theta,\cos\theta\right)$, the frequency becomes simply $\omega_{k}  =N\cos\theta$ and
the interaction coefficients are given by
\begin{align}
\Gamma_{pqk} & =N\frac{1}{\sqrt{2}^{3}}\left(\sin\theta_{q}+\sin\theta_{p}+\sin\theta_{k}\right)\left(K_{p}+K_{q}+K_{k}\right).
\end{align}
The kinetic equation (\ref{eq:pos kin}) has the isotropic equilibrium solution $m_{k}^{-1}\left(K,\theta\right)=1+T^{-1} K$.
Due to the existence of resonantly interacting triads of the positive
branch, this kinetic equation should have cascade solutions
with constant fluxes (see Figure 1 for an illustration of a resonant manifold of eastward propagating waves).
We write (\ref{eq:pos kin}) as a continuity equation 
\begin{align}
\dot{m}\left(\mathbf{k}\right)+\text{div}\Pi\left(\mathbf{k}\right)&=0.\label{eq:continuity}
\end{align}
here, similarly to (\ref{eq:full continuity}), $\Pi$ is the two-dimensional flux. Integrating (\ref{eq:continuity}) wrt the angle $\theta$ we write
\begin{align}
	\dot{l_k}+\text{div}\Pi_k&=0\label{eq:continuity radial},
\end{align}
where $\Pi_{k}=K\int_{0}^{2\pi}d\theta\Pi\cdot\hat{K}$ and $l_{k}=\int_{0}^{2\pi} \! m\left(\mathbf{k}\right)d\theta$ are the isotropic parts of the flux and energy spectrum, respectively. If Fjortoft's argument applies for this system, in a turbulent cascade if the radial flux $\Pi_k$ is not zero, it should be negative. Similarly to the full kinetic equation, a formal solution of (\ref{eq:continuity radial}) is $l_k\propto K^{-3}$.
Constant flux solutions 
are studied in a following paper. 

\section*{{\large{}Appendix I: Off diagonal second order anomalous correlators}}

In writing a kinetic theory for the averaged energy density $n_{\alpha}=\left\langle z_{\alpha}z_{\alpha}^{*}\right\rangle _{\rho}$
one needs to make sure that other second order correlators remain
zero, otherwise these correlators should be added to the kinetic equation
(\ref{eq:kin}) or at least monitored. Such off diagonal correlators
sometimes referred to as anomalous correlators \cite{anom}. The relevant
off diagonal correlator for (\ref{eq:z_epsilon_a}) is 
$\left\langle z_{+}z_{-}^{*}\right\rangle $. Even though initial conditions
are Gaussian the time derivative $\frac{d}{dt}\left\langle z_{+}z_{-}^{*}\right\rangle \neq0$
is not zero. We show however that it is zero in the weak sense of
distributions; that is, rapidly fluctuating around zero in the kinetic
limit $t\omega\rightarrow\infty.$ Writing the correlator in terms of the Fourier expansion of the stream function and elevation:
\begin{align}
\left\langle z_{+}\left(\mathbf{k}\right)z_{-}^{*}\left(\mathbf{k}\right)\right\rangle &=\frac{1}{2}\left\langle \left(k^{2}+m^{2}\right)\hat{A}\left(\mathbf{k}\right)\hat{A}^{*}\left(\mathbf{k}\right)-N^{2}\hat{\zeta}\left(\mathbf{k}\right)\hat{\zeta}^{*}\left(\mathbf{k}\right)\right\rangle +\sqrt{k^{2}+m^{2}}N\left\langle \hat{A}\left(\mathbf{k}\right)\hat{\zeta}^{*}\left(\mathbf{k}\right)-\hat{\zeta}\left(\mathbf{k}\right)\hat{A}^{*}\left(\mathbf{k}\right)\right\rangle,
\end{align}
we see that the physical interpretation of the case where the two brackets fluctuate each around zero in the equation above is that the kinetic energy stored at each wave number $\mathbf{k}$ equals to the potential energy stored at this wave number and that the PM stored at $\mathbf{k}$ equals to the PM stored at $\mathbf{-k}$.

Let us compute the time derivative of the product $z_{\left(+,\mathbf{k}\right)}z_{\left(-,\mathbf{k}\right)}^{*}$:
\begin{align}
\frac{d}{dt}\left(z_{\left(+,\mathbf{k}\right)}z_{\left(-,\mathbf{k}\right)}^{*}\right) & =\frac{1}{2}\epsilon \int\! d\gamma d\beta \left( \sigma_{\left(+,\mathbf{k}\right)}^{\beta\gamma}z_{\beta}^{*}z_{\gamma}^{*}z_{\left(-,\mathbf{k}\right)}^{*}e^{i\left(\omega_{\gamma}+\omega_{\beta}+\omega_{\left(+,\mathbf{k}\right)}\right)t}+\sigma_{\left(-,\mathbf{k}\right)}^{\beta\gamma}z_{\left(+,\mathbf{k}\right)}z_{\beta}z_{\gamma}e^{-i\left(\omega_{\gamma}+\omega_{\beta}+\omega_{\left(-,\mathbf{k}\right)}\right)t}\right)
\end{align}
to write the kinetic equation for $\left\langle z_{\left(+,\mathbf{k}\right)}z_{\left(-,\mathbf{k}\right)}^{*}\right\rangle$ we
take the Laplace transform and integrate by parts once

\begin{align}
s\mathcal{L}\left(z_{\left(+,\mathbf{k}\right)}z_{\left(-,\mathbf{k}\right)}^{*}\right)-\left(z_{\left(+,\mathbf{k}\right)}z_{\left(-,\mathbf{k}\right)}^{*}\right)\mid_{t=0} & =\frac{1}{2}\epsilon J\left(3\right)\nonumber \\
 & -\frac{1}{2}\epsilon\int_{0}^{\infty}\int\! d\gamma d\beta \sigma_{\left(+,\mathbf{k}\right)}^{\beta\gamma}\frac{d}{dt}\left(z_{\beta}^{*}z_{\gamma}^{*}z_{\left(-,\mathbf{k}\right)}^{*}\right)\frac{e^{i\left(\omega_{\gamma}+\omega_{\beta}+\omega_{\left(+,\mathbf{k}\right)}\right)t}}{i\left(\omega_{\gamma}+\omega_{\beta}+\omega_{\left(+,\mathbf{k}\right)}\right)-s}e^{-st}dt\label{eq:anom1}\\
 & -\frac{1}{2}\epsilon\int_{0}^{\infty}\int\! d\gamma d\beta \sigma_{\left(-,\mathbf{k}\right)}^{\beta\gamma}\frac{d}{dt}\left(z_{\left(+,\mathbf{k}\right)}z_{\beta}z_{\gamma}\right)\frac{e^{-i\left(\omega_{\gamma}+\omega_{\beta}+\omega_{\left(-,\mathbf{k}\right)}\right)t}}{i\left(\omega_{\gamma}+\omega_{\beta}+\omega_{\left(+,\mathbf{k}\right)}\right)-s}e^{-st}dt,
\end{align}
where $J\left(3\right)$ stands for third order polynomials which
do not contribute to the kinetic equation and $\mathcal{L}$ for the
Laplace transform. Let us consider term on the RHS of (\ref{eq:pos kin})
\begin{align*}
 & \frac{1}{2}\epsilon\int_{0}^{\infty}\frac{d}{dt}\left(z_{\beta}^{*}z_{\gamma}^{*}z_{\left(-,\mathbf{k}\right)}^{*}\right)\frac{e^{i\left(\omega_{\gamma}+\omega_{\beta}+\omega_{\left(+,\mathbf{k}\right)}\right)t}}{i\left(\omega_{\gamma}+\omega_{\beta}+\omega_{\left(+,\mathbf{k}\right)}\right)-s}e^{-st}dt\\
 & =-\left(\frac{1}{2}\epsilon\right)^{2}\int_{0}^{\infty}\left( \int_{0}^{\infty}\int\! d\beta' d\gamma'\sigma_{\beta}^{\beta'\gamma'}z_{\beta'}z_{\gamma'}z_{\gamma}^{*}z_{\left(-,\mathbf{k}\right)}^{*}\right)\frac{d}{dt}\int_{0}^{t}\frac{e^{i\left(\omega_{\gamma}+\omega_{\beta}+\omega_{\left(+,\mathbf{k}\right)}\right)t'}}{i\left(\omega_{\gamma}+\omega_{\beta}+\omega_{\left(+,\mathbf{k}\right)}\right)-s}e^{-i\left(\omega_{\gamma'}+\omega_{\beta'}+\omega_{\beta}\right)t'}e^{-st'}dt+...
\end{align*}
Integrate by parts once more and take the average with respect to
the Gaussian initial distribution we arrive at
\begin{align}
\mathcal{L}\left(z_{\left(+,\mathbf{k}\right)}z_{\left(-,\mathbf{k}\right)}^{*}\right)-\frac{\left(z_{\left(+,\mathbf{k}\right)}z_{\left(-,\mathbf{k}\right)}^{*}\right)\mid_{t=0}}{s} & =\frac{1}{2}\epsilon^{2}\int\! d\beta d\gamma \sigma_{\left(+,\mathbf{k}\right)}^{\beta\gamma}\left(\sigma_{\beta}^{\gamma\left(-,\mathbf{k}\right)}n_{\gamma}n_{\left(-,\mathbf{k}\right)}+\text{permutations}\right)\cross\\
 & \left(\frac{1}{s\left(i\left(\omega_{\gamma}+\omega_{\beta}+\omega_{\left(+,\mathbf{k}\right)}\right)-s\right)\left(2i\omega_{\left(+,\mathbf{k}\right)}-s\right)}\right)+...+O\left(\epsilon^{3}\right).
\end{align}
Considering all the other contributions, take the inverse Laplace
transform and take the time derivative we obtain the equation for
the off-diagonal correlator 
\begin{align}
\frac{d}{dt}\left\langle z_{\left(+,\mathbf{k}\right)}z_{\left(-,\mathbf{k}\right)}^{*}\right\rangle  & =\frac{1}{2}\epsilon^{2}\int\! d\beta d\gamma \sigma_{\left(+,\mathbf{k}\right)}^{\beta\gamma}\left(\sigma_{\beta}^{\left(-,\mathbf{k}\right)\gamma}n_{\gamma}n_{\left(-,\mathbf{k}\right)}+\sigma_{\gamma}^{\beta\left(-,\mathbf{k}\right)}n_{\beta}n_{\left(-,\mathbf{k}\right)}+\sigma_{\left(-,\mathbf{k}\right)}^{\beta\gamma}n_{\beta}n_{\gamma}\right)\cross\\
 & e^{i\left(\omega_{\gamma}+\omega_{\beta}+\omega_{\left(+,\mathbf{k}\right)}\right)t}\int_{0}^{t}e^{-i\left(\omega_{\gamma}+\omega_{\left(-,\mathbf{k}\right)}+\omega_{\beta}\right)t'}dt'\\
 & +\frac{1}{2}\epsilon^{2}\int\! d\beta d\gamma\sigma_{\left(-,\mathbf{k}\right)}^{\beta\gamma}\left(\sigma_{\beta}^{\left(+,\mathbf{k}\right)\gamma}n_{\gamma}n_{\left(+,\mathbf{k}\right)}+\sigma_{\gamma}^{\beta\left(+,\mathbf{k}\right)}n_{\beta}n_{\left(+,\mathbf{k}\right)}+\sigma_{\left(+,\mathbf{k}\right)}^{\beta\gamma}n_{\beta}n_{\gamma}\right)\cross\nonumber \\
 & e^{-i\left(\omega_{\gamma}+\omega_{\beta}+\omega_{\left(-,\mathbf{k}\right)}\right)t}\int_{0}^{t}e^{i\left(\omega_{\gamma}+\omega_{\left(+,\mathbf{k}\right)}+\omega_{\beta}\right)t'}dt'.
\end{align}
Let us write explicitly the imaginary and real parts of one of the
oscillating terms on the right 

\begin{align}
\Im\left(e^{i\left(\omega_{\gamma}+\omega_{\beta}+\omega_{\left(+,\mathbf{k}\right)}\right)t}\int_{0}^{t}e^{-i\left(\omega_{\gamma}+\omega_{\left(-,\mathbf{k}\right)}+\omega_{\beta}\right)t}\right) & =i\cos\left(\omega_{\gamma}+\omega_{\beta}+\omega_{\left(+,\mathbf{k}\right)}\right)t\frac{1-\cos\left(\omega_{\gamma}+\omega_{\left(-,\mathbf{k}\right)}+\omega_{\beta}\right)t}{\left(\omega_{\gamma}+\omega_{\left(-,\mathbf{k}\right)}+\omega_{\beta}\right)}\\
 & -i\sin\left(\omega_{\gamma}+\omega_{\beta}+\omega_{\left(+,\mathbf{k}\right)}\right)t\frac{\sin\left(\omega_{\gamma}+\omega_{\left(-,\mathbf{k}\right)}+\omega_{\beta}\right)t}{\left(\omega_{\gamma}+\omega_{\left(-,\mathbf{k}\right)}+\omega_{\beta}\right)},\nonumber 
\end{align}

\begin{align}
\Re\left(e^{i\left(\omega_{\gamma}+\omega_{\beta}+\omega_{\left(+,\mathbf{k}\right)}\right)t}\int_{0}^{t}e^{-i\left(\omega_{\gamma}+\omega_{\left(-,\mathbf{k}\right)}+\omega_{\beta}\right)t}\right) & =-\sin\left(\omega_{\gamma}+\omega_{\beta}+\omega_{\left(+,\mathbf{k}\right)}\right)t\frac{1-\cos\left(\omega_{\gamma}+\omega_{\left(-,\mathbf{k}\right)}+\omega_{\beta}\right)t}{\left(\omega_{\gamma}+\omega_{\left(-,\mathbf{k}\right)}+\omega_{\beta}\right)}\\
 & -\cos\left(\omega_{\gamma}+\omega_{\beta}+\omega_{\left(+,\mathbf{k}\right)}\right)t\frac{\sin\left(\omega_{\gamma}+\omega_{\left(-,\mathbf{k}\right)}+\omega_{\beta}\right)t}{\left(\omega_{\gamma}+\omega_{\left(-,\mathbf{k}\right)}+\omega_{\beta}\right)}.\nonumber 
\end{align}
We will show that in the kinetic limit $\omega t\rightarrow\infty$
$f=\sin\left(\omega_{\gamma}+\omega_{\beta}+\omega_{\left(+,\mathbf{k}\right)}\right)t\frac{\sin\left(\omega_{\gamma}+\omega_{\left(-,\mathbf{k}\right)}+\omega_{\beta}\right)t}{\left(\omega_{\gamma}+\omega_{\left(-,\mathbf{k}\right)}+\omega_{\beta}\right)}$
converges weakly to $0$, this similarly follows for the rest of the
terms. 

First let us quickly show that both functions $\sin\left(xs\right),t\sin\left(xs\right)$
converge weakly to $0$ as $s\rightarrow\infty$: Let $\phi$ be a
test function, then using integration by parts
\begin{align}
\lim_{t\rightarrow\infty}\int_{-\infty}^{\infty}\sin\left(xt\right)\phi\left(x\right)dx & =\lim_{t\rightarrow\infty}\int\frac{d}{dx}\left(-\frac{\cos\left(xt\right)}{t}\right)\phi\left(x\right)dx=\lim_{t\rightarrow\infty}\frac{1}{t}\int\cos\left(xt\right)\phi_{x}\left(x\right)dx=\lim_{t\rightarrow\infty}\frac{C}{t}=0\label{eq:sin}
\end{align}
and
\begin{align}
\lim_{t\rightarrow\infty}\int_{-\infty}^{\infty}t\sin\left(xt\right)\phi\left(x\right)dx & =\lim_{t\rightarrow\infty}\int\frac{d}{dx}\left(-\cos\left(xt\right)\right)\phi\left(x\right)dx=\lim_{t\rightarrow\infty}\int\frac{d}{dx}\left(\frac{\sin\left(xt\right)}{t}\right)\phi_{x}\left(x\right)dx\nonumber \\
 & =-\lim_{t\rightarrow\infty}\int\frac{\sin\left(xt\right)}{t}\phi_{xx}\left(x\right)dx=\lim_{t\rightarrow\infty}\frac{C}{t}=0.\label{eq:tsin}
\end{align}
Let $\Gamma\left(y\right)$ be a bounded domain , $g\left(y\right)$
a nice differential function and $y_{0}\in\left[-1,1\right]$ s.t
$\left(y-y_{0}\right)>0$ then 
\begin{align*}
\int_{\Gamma}dy\sin\left(\left(y-y_{0}\right)t\right)\frac{\sin\left(yt\right)}{y}g\left(y\right) & =\int dy\sin\left(\left(y-y_{0}\right)t\right)\frac{d}{dy}\int_{c}^{y}\frac{\sin\left(y't\right)}{y'}dy'g\left(y\right)\\
 & =-\int dy\frac{d}{dy}\left(\sin\left(\left(y-y_{0}\right)t\right)g\left(y\right)\right)\int_{c}^{y}\frac{\sin\left(y't\right)}{y'}dy'\\
 & =-\int dy \left(-t\cos\left(\left(y-y_{0}\right)t\right)g\left(y\right)+\sin\left(\left(y-y_{0}\right)t\right)g_{y}\left(y\right)\right)\int_{c}^{y}\frac{\sin\left(Ny't\right)}{y'}dy'
\end{align*}
taking the limit $t\rightarrow\infty$, using what we showed in (\ref{eq:sin},\ref{eq:tsin})
and $\left|\Theta\left(y\right)\right|=\left|\lim_{t\rightarrow\infty}\int_{c}^{y}\frac{\sin\left(Ny't\right)}{y'}dy'\right|<2$
we obtain 
\begin{align}
\lim_{t\rightarrow\infty}\int_{\Gamma}dy\sin\left(\left(y-y_{0}\right)t\right)\frac{\sin\left(yt\right)}{y}g\left(y\right) & \sim\int dy\lim_{t\rightarrow\infty}\left(-t\cos\left(\left(y-y_{0}\right)t\right)g\left(y\right)+\sin\left(\left(y-y_{0}\right)t\right)g_{y}\left(y\right)\right)2=0.\label{eq:sinsin}
\end{align}
Finally, we show that the integration of the collision kernel can
be brought to a form similar to (\ref{eq:sinsin}). Consider the integral
\begin{align}
\mathcal{I}_{\pm}= & \int K_{p}dK_{p}\int K_{q}dK_{q}\delta\left(\mathbf{k}\right)\int d\theta_{p}\int d\theta_{q}\sin\left(N\left(\cos\theta_{p}+\cos\theta_{q}-\cos\theta_{k}\right)t\right)\frac{\sin\left(N\left(\cos\theta_{p}+\cos\theta_{q}+\cos\theta_{k}\right)t\right)}{\left(\cos\theta_{p}+\cos\theta_{q}+\cos\theta_{k}\right)}\mathcal{K}\label{eq:col pm}
\end{align}
where $\mathcal{K}$ stands for terms in the collision kernel. Let
us change to the variables
\begin{align}
x & =\cos\theta_{p}-\cos\theta_{q}\\
y & =\cos\theta_{p}+\cos\theta_{q}+\cos\theta_{k}
\end{align}
Note that away from the resonant condition, $y=0$, (\ref{eq:col pm})
is zero in the limit $\omega t\rightarrow\infty$. The Jacobian
determinant of the coordinate transformation is $\det J=2\sin\theta_{p}\sin\theta_{q}$,
which is positive in the vicinity of $y=0$; take $\Gamma$ to be
a small vicinity of $y=0$ s.t $\det J\mid_{\Gamma}>0$ in the kinetic
time limit we are we are left with integral of the form 
\begin{align}
\lim_{\omega t\rightarrow\infty}\int K_{p}dK_{p}\int K_{q}dK_{q}\delta\left(\mathbf{k}\right)\int dx\int_{\Gamma}dy\det J\sin\left(N\left(y-2\cos\theta_{k}\right)t\right)\frac{\sin\left(Nyt\right)}{y}\mathcal{K}\left(x,y\right) & =0
\end{align}
this integral vanishes in the limit due to (\ref{eq:sinsin}).

\section*{{\large{}Appendix II: Isotropic part of constant flux solutions of the kinetic equation}}

Consider the kinetic equation (\ref{eq:pos kin}) for eastward propagating waves,
\begin{align}
\dot{m}_{k} & =\epsilon^{2}\pi\int d\mathbf{q}\int d\mathbf{p}\omega_{k}\Gamma_{pqk}^{2}m_{k}m_{p}m_{q}\left(\omega_{q}m_{q}^{-1}+\omega_{p}m_{p}^{-1}+\omega_{k}m_{k}^{-1}\right)\delta\left(\omega_{p,q,k}\right)\delta\left(\mathbf{k}+\mathbf{q}+\mathbf{p}\right).\label{eq:app kin pos}
\end{align}
The interaction coefficients, frequencies, and resonant manifold parameterization 
are given in terms of separable functions in angles and wave number
amplitudes. The interaction coefficient is a homogeneous function wrt
the wave number amplitudes. So it makes sense to look for solutions
in the separable form
\begin{align}
m\left(\mathbf{k}\right) & =\frac{f\left(\theta\right)}{K^{w}}.
\end{align}
We will show that $w=\delta+d=3$ has the KZ scaling, where $\delta=1$
is the homogeneity degree of $\Gamma$ wrt to the wave vector amplitudes
and $d=2$ is the dimension. To determine $w$ let us integrate (\ref{eq:app kin pos})
wrt to the angle and write the isotropic collision kernel as three
identical copies 

\begin{align}
\dot{l}_{k}=\int_{0}^{2\pi}d\theta_{k}\dot{m}\left(k\right) & =\epsilon^{2}\pi\int dK_{p}\int dK_{q}\left(\frac{1}{3}\mathcal{I}_{k}+\frac{1}{3}\mathcal{I}_{k}+\frac{1}{3}\mathcal{I}_{k}\right),\label{eq:app kin pos radial}
\end{align}
where $\mathcal{I}_{k}=K_{p}K_{q}\int_{0}^{2\pi}d\theta_{q}\int_{0}^{2\pi}d\theta_{p}\int_{0}^{2\pi}d\theta_{k}\omega_{k}\Gamma_{pqk}^{2}n_{k}n_{p}n_{q}\left(\omega_{q}n_{q}^{-1}+\omega_{p}n_{p}^{-1}+\omega_{k}n_{k}^{-1}\right)\delta\left(\omega_{p,q,k}\right)\delta\left(\mathbf{k}+\mathbf{q}+\mathbf{p}\right)$.
Using a similar approach, one can show that the isotropic part scaling of the
constant flux solution to the full kinetic equation in the case that
PM is symmetrically distributed $n_{+}\left(k\right)=n_{-}\left(k\right)$
has the same scaling. 
Let us transfrom the second term on the RHS of (\ref{eq:app kin pos radial})
using Zakharov transformation 
of wave number amplitudes and permutation
of the angles:
\begin{align}
K_{q} & =\frac{K^{2}}{K'_{q}},K_{p}=K\frac{K_{p}'}{K_{q}'},\\
\theta_{q}\rightarrow\theta_{k}, & \theta_{k}\rightarrow\theta_{q'},\theta_{p}=\theta_{p'},
\end{align}
and a similar transformation of the third term
\begin{align}
K_{p} & =\frac{K^{2}}{K'_{p}},K_{q}=K\frac{K_{q}'}{K_{p}'},\\
\theta_{p}\rightarrow\theta_{k}, & \theta_{k}\rightarrow\theta_{p'},\theta_{q}=\theta_{q'}.
\end{align}
These transformations transform the collision kernal to itself times
a factor. That brings (\ref{eq:app kin pos radial}) to the following form
\begin{align}
\dot{l}_{k} & =\frac{1}{3}\epsilon^{2}\pi\int dK_{p}\int dK_{q}\left[\omega_{k}+\omega_{q}\left(\frac{K}{K_{q}}\right)^{y}+\omega_{p}\left(\frac{K}{K_{p}}\right)^{y}\right]\mathcal{I}_{k},
\end{align}
where $y=-2w+6$. If $y=0$ the rectangular brackets are proportional
to the resonant condition and vanish. So $w=\delta+d=3$
is a formal solution of the isotropic kinetic equation (\ref{eq:app kin pos radial}).
Its locality and relevance as a solution of (\ref{eq:app kin pos})
is discussed in a following paper.

\end{document}